\documentclass[12pt]{iopart}
\usepackage{graphicx}
\usepackage{epsfig}
\expandafter\let\csname equation*\endcsname\relax
\expandafter\let\csname endequation*\endcsname\relax
\usepackage{amsmath}
\usepackage{iopams}
\usepackage{tikz}

\newcommand{\bra}[1]{\ensuremath{\langle{#1}|\,}}
\newcommand{\ket}[1]{\ensuremath{\,|{#1}\rangle}}
\newcommand{\braket}[1]{\ensuremath{\langle{#1}\rangle}}

\begin{document}

\title[Distribution of waiting times in Lindblad dynamics ]{Distribution of waiting times  between superoperator quantum jumps in Lindblad dynamics}

\author{Daniel S. Kosov}
\address{College of Science and Engineering, James Cook University, Townsville, QLD, 4811, Australia }


\begin{abstract}
Time-evolution of open, dissipative quantum system is a stochastic process that consists of  a  series of  quantum jumps that take place at  random times.  Between quantum jumps  quantum system idles for some time in a particular quantum state. 
Based on superfermion/superboson formalism  and general Kossakowski-Lindblad master equation for an open quantum system we develop a method to compute distribution  of waiting times between quantum jumps.
We illustrate the use of the theory by computing waiting time distribution for model Fermi-system in thermodynamic equilibrium.
\end{abstract}

\date{\today}
\maketitle

The waiting time distribution is a powerful theoretical tool to study stochastic processes \cite{vanKampen}. The dynamics of open quantum system coupled to macroscopic environment is inherently noisy due to quantum jumps, which mixes pure states of the system. Waiting time distribution seems to be natural choice to be used to obtain understanding of the physical mechanisms of quantum irreversibility and randomness. Despite the fact that the concepts of waiting time and quantum mechanics were brought together more than 30 years ago to describe photon counting experiments in optics \cite{Scully1969,Srinivas2010,Vyas1988}, there were almost no interest in pursuing this direction until very recent years, when we started to see a flurry of activity in the use of waiting time distribution to study single electron transport in nanoscale quantum conductors  \cite{brandes08,harbola15,buttiker12,sothmann14,rudge16a}.  In this paper, we develop a general formalism for computing distribution of waiting times between quantum jumps for open quantum systems of bosons and fermions. The approach is based on Kossakowski-Lindblad  (KL) master equation and superfermion/superboson representation of quantum jumps.

Let us consider an open  quantum system described by 
KL master equation\cite{Kossakowski1972,Lindblad1976}:
\begin{equation}
 \dot \rho (t) = -i [H, \rho(t)] + \sum^N_{\alpha=1} \left( 2 A_\alpha \rho(t) A^\dag_\alpha-  [A^\dag_\alpha  A_\alpha, \rho(t)]_+ \right),
\label{kl}
\end{equation}
where $H$ is the Hamiltonian of the system,  $\rho(t)$ is the density matrix,  and $A_{\alpha}$ is a set of  $N$ generally non-Hermitian Lindblad operators that
represent the influence of the environment on the system $[A,B]_\pm = AB \pm BA$  denotes commutators/anticommutators throughout the paper. The Lindblad master equation is the most general master equation which can be derived under the requirements that all probabilities are real and nonnegative, $\rho(t)$ is always normalized, and $\rho(t)$ can be obtained from initial density matrix by linear map \cite{Kossakowski1972,Lindblad1976}.
noise from the quantum jumps. In Fock space, the Lindblad operators  can be generally written as functions of particle creation $a^\dag$ and annihilation  $a$ operators 
\begin{equation}
A_\alpha = A_\alpha(a^\dag, a).
\end{equation}

The Fock space of the system under consideration is spanned by the complete orthonormal set of basis vectors $\ket{m}=\ket{m_1,m_2,\ldots}$, which are eigenvectors of the particle number operator:
\begin{equation}\label{compl}
  a^\dag_k a_k\ket{m}=m_k\ket{m},~~\sum_m \ket{m}\bra{m} = I,~~~\langle n | m \rangle = \delta_{nm}.
\end{equation}
The operators in the Fock space form themselves a linear  vector space called the Liouville-Fock (or super-Fock) space. 
The set of vectors $\ket{mn}\equiv\ket{m}\bra{n}$ form a  orthonormal basis in the Liouville-Fock space. Thus, every operator $A = \sum_{mn} A_{mn}\ket{m}\bra{n}$ can be considered as a
Liouville-Fock space ket-vector $\ket{A} = \sum_{mn}A_{mn}\ket{mn}$. The adjoint operator $A^\dag$ is represented by the bra-vector~$\bra{A}$.
The identity operator corresponds to ket vector  $\ket{I} = \sum_{m}\ket{mm}$.
The  scalar product in the Liouville-Fock space is defined as  $\braket{A_1|A_2} = \mathrm{Tr}(A_1^\dag A_2)$.
In particular, the scalar product of a vector $\ket{A}$ with $\bra{I}$
is equivalent to the trace operation in the Fock space, $\braket{I|A}=\mathrm{Tr}(A)$.

What is described in the paragraph above is the standard method of superoperators widely used in nonequilibrium quantum statistical mechanics.  The subtleties of the approach is in the necessity to introduce 
superoperator partners  for particle creation and annihilation operators.
 Creation and annihilation superoperators are defined as (below we give a brief introduction, for detailed formalism see~\cite{schmutz78,prosen08,dzhioev11a, dzhioev11b,dzhioev12,dzhioev14,dzhioev15,prosen2010}):
\begin{align}\label{def_superop}
  \hat a_k\ket{mn} \equiv a_k \ket{m}\bra{n},~~  \widetilde a_k\ket{mn} \equiv \underbrace{ i(-1)^\mu}_{\text{only for fermions}}\ket{m}\bra{n} a^\dag_k
  \notag\\
 \hat a^\dag_k\ket{mn} \equiv a^\dag_k \ket{m}\bra{n} , ~~  \widetilde a^\dag_k\ket{mn} \equiv \underbrace{i(-1)^\mu}_{\text{only for fermions}}\ket{m}\bra{n} a_k,
\end{align}
where $\mu = \sum_k(m_k + n_k)$. The the phase factor $i(-1)^\mu$ should be dropped in the definition of boson superoperators. We sometimes call $\widetilde a_k, \widetilde a^\dag_k$
tilde superopreators and  $ \hat a_k,  \hat a^\dag_k$ non-tilde superopreators.
These creation/annihilation superoperators satisfy the standard commutations (for bosons) and anticommutation (for fermions) relations:
\begin{equation}
[\hat a_k, \hat a_l^\dag]_{\pm} = \delta_{kl}, \;\;\; [\widetilde a_k, \widetilde a_l^\dag]_{\pm} = \delta_{kl}
\end{equation}
\begin{equation}
[\hat a_k, \hat a_l]_{\pm} = [\widetilde a_k, \widetilde a_l]_{\pm} =0
\end{equation}
The tilde and non-tilde superoperators always commute (bosons) or anticommute (fermions) with each other:
\begin{equation}
[\hat a_k, \widetilde a^\dag_l]_{\pm} = [\hat a_k, \widetilde a_l]_{\pm} =  [\hat a^\dag_k, \widetilde a^\dag_l]_{\pm}=[\widetilde a_k, \hat a^\dag_l]_{\pm}= 0.
\end{equation}
For any operator $A = A(a^\dag, a)$  we formally have two superoperator partners (non-tilde and tilde superoperator)
\begin{equation}\label{superA}
  \hat A = A(\hat a^\dag, \hat a),~~ \widetilde A = A^*(\widetilde a^\dag, \widetilde a).
\end{equation}
The connection between non-tilde and tilde superoperators is given by the "tilde conjugation rules"~\cite{dzhioev11a,dzhioev14}:
\begin{align}\label{TC_rules}
 &(c_1 \hat A_1 + c_2 \hat A_2)\widetilde{} = c_1^* \widetilde{A}_1 + c^*_2 \widetilde {A}_2,
 \notag\\
 &(\hat A_1\hat A_2)\widetilde{}=\widetilde{A}_1\widetilde{A}_2, ~~ (\widetilde A)\widetilde{} = \hat A,~~(\hat A^\dag)\widetilde{} = (\widetilde A)^\dag.
\end{align}
By means of superoperators an arbitrary  Liouville-Fock space vector can be represented as
\begin{equation}\label{ketA}
  \ket{A} = \hat A\ket{I} = \sigma_A \widetilde A^\dag\ket{I}.
\end{equation}
Hereinafter, the phase 
\begin{eqnarray}
\sigma_A =
\left\{ \begin{array}{c}
-i, \text{  if  } A \text{   is  fermionic operator}, \\
+1, \text{  if  } A \text{   is  bosonic operator}. \end{array} \right.
\end{eqnarray}
Considering the adjoint of~\eqref{ketA} and taking  $\hat A$ as $\hat a~\mathrm{and}~\hat a^\dag$  we get
\begin{equation}\label{bra_vac}
  \bra{I}(\hat a^\dag -  \sigma^*_a \widetilde a)=\bra{I}(\widetilde a^\dag -  \sigma_a \hat a)=0.
\end{equation}
Therefore  superoperator $\hat b^\dag = \hat a^\dag -  \sigma^*_a \widetilde a$ and its tilde conjugate, $\widetilde b^\dag=\widetilde a^\dag -  \sigma_a \hat a$
annihilate the bra-vector $\bra{I}$.

Having defined creation and annihilation superoperators, we can start to work with KL master equation.
We convert KL master equation to superoperator form by multiplying it from the right on vector $\ket{I}$. The KL master equation becomes
\begin{equation}
\ket{\dot \rho (t)} = -i \mathcal L \ket{ \rho(t)},
\label{kl}
\end{equation}
where the Liouvillian is 
\begin{equation}
-i \mathcal L=-i (\hat H-\widetilde H) \ket{\rho (t)} - \sum^N_{\alpha=1} \left(   \hat A^\dag_\alpha  \hat A_\alpha + \widetilde A^\dag_\alpha  \widetilde A_\alpha - 2 \sigma_{A_\alpha}  \; \hat A_\alpha  \widetilde A_\alpha  \right).
\label{L}
\end{equation}
Combining Lamb shift and system Hamiltonian into one superperator (t.c. is tilde conjugation operation performed by  "tilde conjugation rules" (\ref{TC_rules}))
\begin{equation}
\mathcal L_0= \left(\hat H - i \sum^N_{\alpha=1} \hat A^\dag_\alpha \hat A_\alpha \right) - \text{t.c.}
\end{equation}
and introducing jump superoperators
\begin{equation}
J_\alpha = 2 \sigma_{A_\alpha} \hat A_\alpha  \widetilde A_\alpha,
\end{equation}
we bring  KL master equation to the following form
\begin{equation}
 \ket{\dot \rho (t)} =(-i  \mathcal L_0 + \sum^N_{\alpha=1} J_\alpha ) \ket{ \rho(t)}.
\label{kl}
\end{equation}

Time-evolution of open, dissipative quantum system is a stochastic process that consists of  a  series of  quantum ``jumps" described by superoperators $J_\alpha$ that take place at specific but random times.  Between quantum jumps the dynamics of quantum system is generated by superoperator $-i  \mathcal L_0$. Suppose that we observe 
quantum jump described by superoperator $  J_\alpha   $ at  $t_1$, and then
 we observe the next quantum jump   $ J_\beta$ at some later  time  $t_2$.
The times of quantum jumps, $t_1$ and $t_2$,  are stochastic variables distributed on the positive real axis.
Let us define the following probability distributions:
\\
$P_{\beta \alpha} (t_2, t_1)$  -- the partial joint probability distribution that the quantum system undergoes  a jump $\alpha$  at time  $t_1$
 and quantum jump of type $\beta$ at time   $t_2$,
\\
$p_\alpha (t)$  -- the probability distribution that the system performed quantum jump $\alpha$  in time  $t$,
 \\
 $w_{\beta \alpha}(t_2, t_1)$ -- the conditional probability that the quantum system undergoes  a jump $\alpha$  at time  $t_1$
 and quantum jump of type $\beta$ at time   $t_2$  that there were no other quantum jumps between time $t_1$ and $t_2$.  The quantity   $w_{\beta \alpha}(t_{2},t_{2})$ is called  {\it waiting time distribution} between subsequent quantum jumps.
These three probability distributions  are  connected to each other by the standard relation between joint and conditional probabilities:
\begin{equation}
P_{\beta \alpha} (t_2, t_1) =  w_{\beta \alpha}(t_2, t_1) p_{\alpha}(t_1).
\label{kolmogorov}
\end{equation}
We note that  in a stationary regime (that may be thermodynamic equilibrium or nonequilibrium steady state), $w_{\beta \alpha}(t_2, t_1)$  and  $P_{\beta \alpha}(t_2, t_1)$ depend on relative time only, $\tau = t_2 -t_1$, and $p_{\alpha}(t)$ becomes time-independent.

To develop the formalism associated with the tunnelling time distribution, we let the system evolve 
to time $t_1$ starting from some initial density matrix $\ket{\rho_0}$
\begin{equation}
\ket{\rho(t_1)} = e^{-i \mathcal L t_1} \ket{\rho_0}
\end{equation}
and then we begin to monitor quantum jumps.
First, we define the partial joint probability distribution that  the system performs quantum jump $\alpha$ at some  time  $t_1$,
 idle without quantum jumps for some time $t_2 - t_1$ and then undergoes jump $\beta$ at   $t_2$:
\begin{eqnarray}
P_{\beta \alpha} (t_2,t_1) &&=   \bra{I} \left[2 \sigma_{A_\beta}\;  {\hat A}_{\beta} \widetilde A_{\beta} \right]  e^{ -i\mathcal L_0 (t_2-t_1)}  \left[ 2 \sigma_{A_\alpha}\; {\hat A}_{\alpha} \widetilde A_{\alpha} \right] \ket{ \rho(t_1)} 
\nonumber 
\\
&&=
4 \sigma_{A_\beta} \sigma_{A_\alpha} \; \bra{I}  {\hat A}_{\beta} \widetilde A_{\beta} e^{ -i\mathcal L_0 (t_2-t_1)}  {\hat A}_{\alpha} \widetilde A_{\alpha} \ket{ \rho(t_1)}.
\label{P}
\end{eqnarray}
The probability that the system undergoes jump  $\alpha$ at time $t_1$ is
\begin{equation}
p_{\alpha}(t_1) = 2 \sigma_{A_\alpha} \;  \bra{I}  {\hat A}_{\alpha} \widetilde A_{\alpha} \ket{ \rho(t_1)}.
\label{p}
\end{equation}
Using  equations for probabilities (\ref{P})  and (\ref{p}) along with relation (\ref{kolmogorov}), we obtain the expression for the waiting time distribution 
\begin{equation}
w_{\beta \alpha} (t_2,t_1) = 2 \sigma_{A_\beta} \; \frac{ \bra{I}  {\hat A}_{\beta} \widetilde A_{\beta} e^{ -i\mathcal L_0 (t_2-t_1)} {\hat A}_{\alpha} \widetilde A_{\alpha} \ket{ \rho(t_1)} }{ \bra{I} {\hat A}_{\alpha} \widetilde A_{\alpha} \ket{ \rho(t_1)} } 
\label{w}
\end{equation}
There are $N^2$ physically different waiting time distributions where $N$ is the number of distinct Lindblad operators in KL master equation (\ref{kl}). The waiting time distribution expressed in terms  of Linddlad operators (\ref{w})
is one of the main results of the paper: It is general, applicable to both bosons and fermions and can be readily used for any kind of open quantum system described by KL master equation.

The normalisation of the waiting time distribution deserves special discussion. Over all time $ t_2 \in [t_1,+\infty[$ the probability for a quantum tunnelling event to occur is unity. Therefore, we sum over all possible final quantum jumps $\sum_{\beta}$ and integrate over time $t_2$ to obtain normalisation:
\begin{eqnarray}
\notag
  \sum^N_{\beta=1} \int_{t_1}^{\infty} d t_2 w_{\beta \alpha} (t_2,t_1) =
  \sum^N_{\beta=1} \int_{t_1}^{\infty} d t_2 \; 2 \sigma_{A_\beta} \; \frac{ \bra{I}  {\hat A}_{\beta} \widetilde A_{\beta} e^{ -i \mathcal L_0 (t_2-t_1)} {\hat A}_{\alpha} \widetilde A_{\alpha} \ket{ \rho(t_1)} 
}{\bra{I} {\hat A}_{\alpha} \widetilde A_{\alpha} \ket{ \rho(t_1)}} 
   && 
  \\
  = \int_{t_1}^{\infty} d t_2  \frac{ \bra{I}-i [\mathcal L- \mathcal L_0] e^{ -i \mathcal L_0 (t_2-t_1)} {\hat A}_{\alpha} \widetilde A_{\alpha} \ket{ \rho(t_1)} 
}{\bra{I} {\hat A}_{\alpha} \widetilde A_{\alpha}\ket{ \rho(t_1)} 
} 
\nonumber
\end{eqnarray}
Since $(I| \mathcal L=0$, the normalisation becomes
\begin{equation}
\int_{t_1}^{\infty} d t_2  \frac{ \bra{I}i \mathcal L_0 e^{ -i\mathcal L_0\tau} {\hat A}_{\alpha} \widetilde A_{\alpha} \ket{ \rho(t_1)} 
}{\bra{I} {\hat A}_{\alpha} \widetilde A_{\alpha} \ket{ \rho(t_1)} 
} = \frac{ \bra{I} \mathcal L_0 \mathcal L_0^{-1} {\hat A}_{\alpha} \widetilde A_{\alpha} \ket{ \rho(t_1)} 
}{\bra{I} {\hat A}_{\alpha} \widetilde A_{\alpha} \ket{ \rho(t_1)} 
} =1.
\end{equation}
That means the waiting time distribution defined by (\ref{w}) is automatically properly normalised when all $N$ secondary quantum jumps are included.

Let us now apply this theory to compute distribution of waiting times for a model system.
Consider a Fermi-system which consists of one energy level
 \begin{equation}
    H= \varepsilon a^\dag a.
 \end{equation}
 The system is open and attached to the macroscopic thermal bath. It is described by the standard KL master equation 
 \begin{equation}
 \dot \rho (t) = -i [H, \rho(t)] + \left( 2 A_1 \rho(t) A^\dag_1-  [A^\dag_1  A_1, \rho(t)]_+ + 2 A_2 \rho(t) A^\dag_2-  [A^\dag_2  A_2, \rho(t)]_+  \right),
\label{kl-1f}
\end{equation}
 where the influence of the bath on the systems  is accounted by two Lindblad operators
 \begin{equation}
    A_1=\sqrt{\Gamma (1-f)}a,~~~A_2=\sqrt{\Gamma f}a^\dag.
 \end{equation}
 Here the bath thermodynamic parameters, temperature $T$ and chemical potential $\mu$, enters the KL master equation via Fermi occupation numbers in the Lindblad operators 
 $f= [ 1+ e^{(\varepsilon - \mu)/T}]^{-1} $.

 In superfermion representation KL master equation (\ref{kl-1f}) becomes 
 \begin{equation}
 \ket{\dot \rho (t)} = \left(-i \mathcal L _0 +J_1 + J_2 \right) \ket{ \rho(t)}.
 \end{equation}
Here the superoperator which describes the streaming part of the density matrix evolution  is
\begin{equation}
\mathcal L_0 = E \hat a^\dag \hat a - E^* \widetilde a^\dag \widetilde a - 2 i \Gamma f,
\end{equation}
with $E= \varepsilon - i \Gamma (1- 2 f)$
and  two quantum jump superoperators are
\begin{equation}
J_1= -2i  \Gamma (1-f) \hat a \widetilde a,
\; \; \; \;
J_2 = -2i  \Gamma f \hat a^\dag \widetilde a^\dag.
\end{equation}
Both jump superoperators are associated with particle and energy exchange  between the system and thermal bath 
From the structure of the superoperators we conclude that jump $J_1$ is associated with particle transfer from the system to the bath and $J_2$ describe the reverse processes when the particle is transported to the system from bath.

The total KL  Liovillian
\begin{equation}
-i \mathcal L =  -i \mathcal L_0 + J_1 + J_2
\end{equation}
is a quadratic, non-Hermitian form in terms of superfermion creation/annihilation superoperators. We can diagonalise it by
the following canonical, nonunitary transformation to nonequilibrium quasiparticles (also called Prosen's normal master modes  \cite{prosen08}):
\begin{align}\label{transf2}
&  \hat a=\hat b- if \widetilde b_\dag ,~~~ \widetilde a = \widetilde b  + if  \hat  b_\dag ,
  \\
  \label{transf3}
&\hat a^\dag =(1-f ) \hat b_\dag +i\widetilde b ,~~ ~\widetilde a^\dag =  (1 - f ) \widetilde b_\dag   - i \hat b.
\end{align}
These transformations are canonical, that means theypreserve the anticommutation relation between the fermion creation/annihilation superoperators:
\begin{equation}
[\hat b, \hat b_\dag]_{+} = 1, \;\;\; [\widetilde b, \widetilde b_\dag]_{+} = 1
\end{equation}
\begin{equation}
[\hat b, \hat b]_{+} = [\widetilde b, \widetilde b]_{+} =[\hat b_\dag, \hat b_\dag]_{+} = [\widetilde b_\dag, \widetilde b_\dag]_{+} =0,
\end{equation}
and the tilde and non-tilde superoperators continue to anticommute  with each other. Although, this transformation is canonical, it is not unitary meaning that creation and annihilation operators for nonequilibrium quasiparticles are not related to each other by Hermitian conjugation: $\hat b_{\dagger} \ne (\hat b)^{\dagger}$ and $\widetilde b_{\dagger} \ne (\widetilde b)^{\dagger}$ 
Since,  $\hat b_\dag=\hat a^\dag-i\widetilde a$, the vector $\bra{I}$  is automatically the vacuum
for $\hat b_\dag$ and $\widetilde b_\dag$ operators:
\begin{equation}
\bra{I} \hat b_\dag = \bra{I} \widetilde b_\dag =0.
\end{equation}
 $\mathcal L $, expressed in terms of nonequilibrium quasiparticles,  $\mathcal L $ becomes diagonal:
\begin{equation}
\label{L}
\mathcal  L =
( \varepsilon + i\Gamma)  \hat b_\dag \hat b  - ( \varepsilon - i\Gamma) \widetilde b_\dag \widetilde b.
\end{equation}

Suppose that the system is in thermodynamic equilibrium that means  it is described by the density matrix which is null vector of the full Liovillian
\begin{equation}
\mathcal L \ket{\rho}  =0.
\end{equation}
It also means that the equilibrium density matrix is a vacuum vector for nonequilibrium quaiparticle annihilation operators
\begin{equation}
 \hat b \ket{\rho} =  \widetilde b \ket{\rho} =0.
\end{equation}

The waiting time distribution between quantum jumps is given by (\ref{w}). It 
does not depend on the initial time $t_1$ for stationary density matrix and is determined only on the time delay between two quantum jumps $\tau = t_2 - t_1$.
Let us consider the distribution of waiting time between the two quantum events: first, at some arbitrary time $t_1$ the fermion appears in the system by quantum jump $J_2$, it spends in the systems time $\tau$ and then  it "jumps out" of the system at time $t_2= t_1 + \tau$:
\begin{equation}
w_{12} (\tau) = -2 i \Gamma (1-f)  e^{-2 \Gamma f \tau} 
\frac{ 
\bra{I}  \hat a \widetilde a e^{ -i(E \hat a^\dag \hat a - E^* \widetilde a^\dag \widetilde a)  \tau} \hat a^\dag \widetilde  a^\dag  \ket{ \rho}
}
{
\bra{I} \hat a^\dag \widetilde a^\dag \ket{ \overline\rho}
}.
\label{w2}
\end{equation}
Using the canonical transformation to nonequilibrium quasiparticle creation and annihilation superoperators (\ref{transf2},\ref{transf3})  the matrix elements in (\ref{w2}) can be readily computed:
\begin{equation}
w_{12} (\tau) = 2 \Gamma (1-f)  e^{-2 \Gamma (1-f) \tau}.
\label{w3}
\end{equation}
This  calculations demonstrate that the maintaining thermodynamic equilibrium is a dynamic processes. The distribution of waiting times between two balancing quantum events - the first one is when the energy/particle is brought into the system and the second jump when energy/particle is removed from the system - is Poisson process. The same result was recently obtained by our group based on the rate equation approach  for occupation probabilities\cite{rudge16a},  this agreement  between rate and KL equations demonstrates that the role of quantum coherences  in waiting time distribution completely disappear from the KL master equation in stationary regime.

In conclusion, based on KL master equation and superfermion/superboson  formalism we develop a general theory for calculation of distribution of waiting times between subsequent quantum  jumps.
The proposed method is universal,  works in  both stationary (steady state nonequilibrium or thermal equilibrium) and transient regimes, is applicable to open quantum systems of interacting bosons and fermions as long as they can be described by KL master equation. 
We obtain useful expression for waiting time distribution in terms of Lindblad operators (\ref{w}) which can be readily employed for calculations using KL master equation for any physical systems.
We  illustrated the theory by calculations of waiting time distributions between quantum jumps for model open  equilibrated Fermi-system.

\section*{References}

\bibliographystyle{unsrt}

\end{document}